\RequirePackage{fix-cm}
\documentclass[smallextended]{svjour3}       
\smartqed  
\usepackage{graphicx}
\usepackage{tgtermes}

\usepackage[T1]{fontenc}
\usepackage[latin9]{inputenc}
\usepackage[a4paper]{geometry}
\usepackage{array}
\usepackage{float}
\usepackage{multirow}
\usepackage{amsmath}
\usepackage{amssymb}
\usepackage{graphicx}

\begin{document}

\title{Bayesian Matrix Completion Approach to Causal Inference with Panel Data
}


\author{Masahiro Tanaka
}


\institute{M. Tanaka \at
              Department of Economics, Kanto Gakuen University; Graduate School of Economics, Waseda University\\
200, Fujiagucho, Ohta, Gunma
169-8050 Japan \\
              \email{gspddlnit45@toki.waseda.jp}
}

\date{July 29, 2020}

\maketitle

\begin{abstract}
This study proposes a new Bayesian approach to infer binary treatment
effects. The approach treats counterfactual untreated outcomes as
missing observations and infers them by completing a matrix composed
of realized and potential untreated outcomes using a data augmentation
technique. We also develop a tailored prior that helps in the identification
of parameters and induces the matrix of untreated outcomes to be approximately
low rank. Posterior draws are simulated using a Markov Chain Monte
Carlo sampler. While the proposed approach is similar to synthetic
control methods and other related methods, it has several notable
advantages. First, unlike synthetic control methods, the proposed
approach does not require stringent assumptions. Second, in contrast
to non-Bayesian approaches, the proposed method can quantify uncertainty
about inferences in a straightforward and consistent manner. By means
of a series of simulation studies, we show that our proposal has a
better finite sample performance than that of the existing approaches.
\keywords{Causal inference \and Matrix completion \and Panel data \and Synthetic control
method \and Treatment effect}
\end{abstract}

\section{Introduction}

Program/policy evaluations and comparative case studies using observational
data are pervasive in social and natural sciences and in government
and business practice. In particular, causal inference is an integral
part of social sciences, where randomized experiments are usually
infeasible. For instance, although Abadie et al. \cite{Abadie2015}
analyzed the economic cost of the German reunification in 1990, we
cannot repeat such a political event many times in a controlled fashion.

The primary interest of this study is inference of causal effects
of a treatment, such as average treatment effect and average treatment
effect on treated (ATET). Suppose we have panel data with $J$ units
and $T$ time periods. An outcome of unit $j$ at period $t$ is denoted
by $y_{j,t}\left(s_{j,t}\right)$, where $s_{j,t}=1$ when the unit
is exposed to treatment and $s_{j,t}=0$ otherwise. Let $\mathcal{I}_{1}$
and $\mathcal{I}_{0}$ be sets of indices for treated and untreated
observations, respectively. Then, for instance, ATET is defined as
\[
ATET=\frac{1}{\left|\mathcal{I}_{1}\right|}\sum_{\left(j,t\right)\in\mathcal{I}_{1}}\left(y_{j,t}\left(1\right)-y_{j,t}\left(0\right)\right),
\]
where $\left|\mathcal{A}\right|$ denotes the cardinality of set $\mathcal{A}$.
Inference of causal effects amounts to inference of counterfactual
untreated outcomes $y_{j,t}\left(0\right)$, $\left(j,t\right)\in\mathcal{I}_{1}$,
or the ``potential outcome'' in terms of Neyman\textendash Rubin's
causal model \cite{Imbens2015}. Causal inference
poses a serious challenge to statisticians, and numerous approaches
have been proposed: the difference-in-difference estimator, regression
discontinuity design, matching-based methods, etc.\footnote{See, e.g., \cite{Imbens2015}.}

In this study, we propose a new Bayesian approach for inferring the causal effect of a binary treatment with panel data. We transform a statistical problem of causal inference into a matrix completion problem, an extensively studied issue in machine learning (e.g., \cite{Keshavan2010}). Our approach implements in two steps. First, the potential outcomes are inferred via a Bayesian matrix completion method. Then, a causal effect is inferred based on the posterior draws of the potential outcomes.

We model the sum of a matrix of outcomes using two-component factorization and a matrix of covariate effects. The potential outcomes are treated as missing observations and simulated from the posterior predictive distribution,
\[
\int p\left(y_{j,t}\left(0\right),\left(j,t\right)\in\mathcal{I}_{1}|\mathcal{D},\boldsymbol{\Theta}\right)\textrm{d}\boldsymbol{\Theta},
\]
where $\mathcal{D}$ denotes a set of observations including untreated outcomes $y_{j,t}\left(0\right)$, $\left(j,t\right)\in\mathcal{I}_{0}$ and exogenous covariates and $\boldsymbol{\Theta}$ denotes a set of parameters and random variables to be sampled. The other unknown parameters, such as coefficients on the covariates and the variance of measurement error, are simulated from the conditional posterior distribution. Leaving aside the covariate effects, the proposed approach can be thought as treating inference of the potential outcomes as multiple imputation of a matrix of panel data that is probably rank deficient.

Given the posterior draws of potential outcomes, we can infer a causal effect of interest. For instance, when we have a total of $N_{post}$ posterior draws of potential outcomes, the posterior mean estimate of the ATET is given by
\[
\widehat{ATET}=\frac{1}{N_{post}}\sum_{i=1}^{N_{post}}\frac{1}{\left|\mathcal{I}_{1}\right|}\sum_{\left(j,t\right)\in\mathcal{I}_{1}}\left(y_{j,t}\left(1\right)-y_{j,t}^{\left(i\right)}\left(0\right)\right),
\]
where $y_{j,t}^{\left(i\right)}\left(s_{j,t}\right)$ denotes the $i$th posterior draw of the potential outcome of unit $j$ at period $t$. 

To facilitate this task, we develop a tailored prior that induces the model to be lower rank, adapting a cumulative shrinkage process prior \cite{Legramanti2019}. With this prior specification, there is no need to specify the rank of the outcome matrix, because the priors pushes insignificant columns of one of the factorizations toward zero.

Our Bayesian approach has two notable advantages. First, it can provide
credible intervals in a consistent and straightforward manner, while
the existing non-Bayesian approaches have difficulty quantifying uncertainty.
As hypothesis testing is an essential component of scientific research,
this advantage is a strong reason to use a Bayesian method. Second,
our approach has better finite sample performance than that of the
existing approaches. By means of a series of simulation studies, we
show that our proposal is competitive with the existing approaches
in terms of the precision of the prediction of potential outcomes. 

Three strands of the literature are particularly relevant to this
study. First, the proposed approach is related to a class of synthetic
control methods (SCMs) (e.g., \cite{Abadie2003,Abadie2010}).\footnote{See \cite{Abadieforthcoming} for a recent overview of the literature
on SCMs.} This class of methods is aimed at obtaining ``synthetic'' observations
of untreated outcomes as weighted sums of the outcomes of the control
units. Despite its increasing popularity, the original SCM \cite{Abadie2010}
has two notable shortcomings. First, it imposes a strong assumption
that the weights of synthetic observations are nonnegative and sum
to one. This assumption implies that the treated unit falls in the
convex hull of the control units and that synthetic observations are
positively correlated with the control units, which is not plausible
in many real situations. While some alternative approaches \cite{Doudchenko2017,Kim2019,Amjad2018}
do not require these assumptions, our approach has better finite sample
performance under various data generating processes, as shown in the
simulation studies. Second, the original SCM does not have an effective
method for assessing the uncertainty of the obtained estimates. Abadie
et al. \cite{Abadie2010} conduct a series of placebo studies, but
the approach incurs size distortion \cite{Hahn2017}. Recently, Li
\cite{Liforthcoming} proposes a subsampling method to obtain confidence
intervals for SCMs, but our Bayesian approach can obtain credible
intervals simply as a byproduct of posterior simulation. 

Second, an approach developed by Athey et al. \cite{Athey2018} is
particularly related to our proposal. They also treat potential outcomes
as missing data and estimate them via matrix completion with the nuclear
norm penalty \cite{Mazumder2010}. However, Athey et al.'s \cite{Athey2018}
non-Bayesian approach does not have an estimator for confidence intervals.
Finally, our proposal is conceptually similar to approaches proposed
by \cite{Brodersen2015,Ning2019} in that all of them infer potential
outcomes as missing observations in Bayesian manners. On the other
hand, their approaches rely on the fit of a time-series model, while
our approach exploits the factor structure of panel data. Therefore,
our proposal is better suited for typical panel data covering short
time periods where it is difficult to estimate a time-series model. 

The remainder of this study is structured as follows. In Section 2,
we introduce a new Bayesian approach to causal inference with panel
data and compare it with the existing alternatives. In Section 3,
we illustrate the proposed approach by applying it to simulated and
real data. We conduct a simulation study and show that our proposed
method is competitive with the existing approaches in terms of the
precision of the predictions of potential outcomes. Then, the proposed
approach is applied to the evaluation of the tobacco control program
implemented in California in 1988. The last section concludes the
study.

\section{Proposed Approach}

\subsection{Framework}

An individual outcome is modeled as follows: for $j=1,...,J$; $t=1,...,T$,

\[
y_{j,t}\left(s_{j,t}\right)=\gamma_{j,t}+\boldsymbol{x}_{j,t}^{\top}\boldsymbol{\beta}+u_{j,t},\quad u_{j,t}\sim\mathcal{N}\left(0,\tau^{-1}\right),
\]
where $\gamma_{j,t}$ is a unit- and time-specific intercept, $\boldsymbol{x}_{j,t}$
is an $L$-dimensional vector of covariates that may contain unit-
and/or time-specific effects, $\boldsymbol{\beta}$ is the corresponding
coefficient vector, and $u_{j,t}$ is an error term that is distributed
according to a normal distribution with precision $\tau$. $s_{j,t}$
is a treatment indicator: $s_{j,t}=0$ if $y_{j,t}$ is untreated
and $s_{j,t}=1$ if $y_{j,t}$ is treated. A set of the treatment
indicators is denoted by $\boldsymbol{S}$=$\left\{ s_{j,t}\right\} $.
The covariates $\boldsymbol{x}_{j,t}$ are completely observed for
all the units and periods. 

$\mathcal{I}_{1}$ and $\mathcal{I}_{0}$ are sets of indices for
treated and untreated observations, respectively. $\mathcal{I}=\mathcal{I}_{1}\cup\mathcal{I}_{0}$
denotes a set of all the indices for the observations. Let $\boldsymbol{Y}$
be a $J$-by-$T$ matrix composed of (actually) untreated outcomes,
$y_{j,t}\left(0\right)$, $\left(j,t\right)\in\mathcal{I}_{0}$, and
counterfactual untreated outcomes that are actually treated, $y_{j,t}\left(0\right)$,
$\left(j,t\right)\in\mathcal{I}_{1}$. The latter elements are also
called ``potential outcomes'' \cite{Imbens2015}.
We define sets of observed and unobserved untreated outcomes respectively
as 
\[
\boldsymbol{Y}^{obs}=\left\{ y_{j,t}\left(0\right),\;\left(j,t\right)\in\mathcal{I}_{0}\right\} ,\quad\boldsymbol{Y}^{miss}=\left\{ y_{j,t}\left(0\right),\;\left(j,t\right)\in\mathcal{I}_{1}\right\} .
\]
In this study, we treat $\boldsymbol{Y}^{miss}$ as missing and infer
it as a set of unknown parameters via matrix completion, using a Markov
chain Monte Carlo (MCMC) method. In other words, we transform the
inferential problem into a matrix completion problem, and infer the
potential outcomes by imputing them via data augmentation \cite{Tanner1987}
(or, more generally, Gibbs sampling). 

Let $\boldsymbol{X}=\left\{ \boldsymbol{x}_{j,t}\right\} $ be a set
of covariates. Define a $J$-by-$T$ matrix of the covariate effects
as $\boldsymbol{\Xi}=\left(\xi_{j,t}\right)$ with $\xi_{j,t}=\boldsymbol{x}_{j,t}^{\top}\boldsymbol{\beta}$.
The model can be posed in a matrix representation as

\[
\boldsymbol{Y}=\boldsymbol{\Theta}+\boldsymbol{\Xi}+\boldsymbol{U},
\]
where $\boldsymbol{U}=\left(u_{j,t}\right)$ is a $J$-by-$T$ matrix
of the error terms. 

We have several assumptions in the model. First, we make the standard
stable unit treatment value assumption (STUVA) \cite{Imbens2015}:
there is no interference between units, there is a single type of
treatment, and each unit has two potential outcomes. In addition,
the assignment mechanism is assumed to be unconfounded, i.e., ignorable,
conditional on the covariates $\boldsymbol{X}$:
\[
p\left(\boldsymbol{S}|\boldsymbol{Y}^{obs},\boldsymbol{Y}^{miss},\boldsymbol{X}\right)=p\left(\boldsymbol{S}|\boldsymbol{X}\right).
\]
In contrast to the difference-in-difference estimator, the treated
and untreated units are not supposed to have parallel trends in outcome. 

A matrix of untreated outcomes $\boldsymbol{Y}$ can be structured
flexibly. For instance, when only the $J$th unit is affected by the
treatment for the last $T_{1}$ periods as in the standard synthetic
control method (SCM) \cite{Abadie2003,Abadie2010}, $\boldsymbol{Y}$
is specified as 

\[
\boldsymbol{Y}=\left(\begin{array}{cccccc}
y_{1,1}\left(0\right) & \cdots & y_{1,T_{0}}\left(0\right) & y_{1,T_{0}+1}\left(0\right) & \cdots & y_{1,T}\left(0\right)\\
\vdots &  & \vdots & \vdots &  & \vdots\\
y_{J-1,1}\left(0\right) & \cdots & y_{J-1,T_{0}}\left(0\right) & y_{J-1,T_{0}+1}\left(0\right) & \cdots & y_{J-1,T}\left(0\right)\\
y_{J,1}\left(0\right) & \cdots & y_{J,T_{0}}\left(0\right) & \surd & \cdots & \surd
\end{array}\right),
\]
where $\surd$ denotes a missing entry. It is possible to allow more
than one treated unit:
\[
\boldsymbol{Y}=\left(\begin{array}{cccccc}
y_{1,1}\left(0\right) & \cdots & y_{1,T_{0}}\left(0\right) & y_{1,T_{0+1}}\left(0\right) & \cdots & y_{1,T_{0}+1}\left(0\right)\\
\vdots &  & \vdots & \vdots &  & \vdots\\
y_{J_{0},1}\left(0\right) & \cdots & y_{J_{0},T_{0}}\left(0\right) & y_{J_{0},T_{0+1}}\left(0\right) & \cdots & y_{J_{0},T_{0}+1}\left(0\right)\\
y_{J_{0}+1,1}\left(0\right) & \cdots & y_{J_{0}+1,T_{0}}\left(0\right) & \surd & \cdots & \surd\\
\vdots &  &  & \vdots &  & \vdots\\
y_{J,1}\left(0\right) & \cdots & y_{J,T_{0}}\left(0\right) & \surd & \cdots & \surd
\end{array}\right).
\]
Furthermore, it is possible to handle a more complex structure: 

\[
\boldsymbol{Y}=\left(\begin{array}{cccccccc}
\surd & \cdots & \surd & y_{1,t^{\prime}}\left(0\right) & \cdots & \cdots & \cdots & y_{1,T}\left(0\right)\\
\cdots & \cdots & \cdots & \cdots & \cdots & \cdots & \cdots & \vdots\\
y_{j^{\sharp},1}\left(0\right) & \cdots & y_{j^{\sharp},t^{\prime}+1}\left(0\right) & \surd & y_{j^{\sharp},t^{\prime}+1}\left(0\right) & \cdots & \cdots & y_{j^{\sharp},T}\left(0\right)\\
\cdots & \cdots & \cdots & \cdots & \cdots & \cdots & \cdots & \cdots\\
y_{j^{\flat},1}\left(0\right) & \cdots & y_{j^{\flat},t^{\prime}+1}\left(0\right) & \surd & \surd & y_{j^{\flat},t^{\prime}+2}\left(0\right) & \cdots & y_{j^{\flat},T}\left(0\right)\\
\vdots & \cdots & \cdots & \cdots & \cdots & \cdots & \cdots & \vdots\\
y_{J,1}\left(0\right) & \cdots & \cdots & \cdots & \cdots & \cdots & \cdots & y_{J,T}\left(0\right)
\end{array}\right).
\]

From a theoretical perspective, Bayesian inference of $\boldsymbol{Y}^{miss}$
is specified as follows. Let $\boldsymbol{\Theta}=\left\{ \boldsymbol{\Gamma},\boldsymbol{\beta},\tau\right\} $
denote the set of all unknown parameters. If $\boldsymbol{Y}^{obs}$
, $\boldsymbol{Y}^{miss}$, and $\boldsymbol{S}$ are given, the complete-data
likelihood is represented as

\begin{eqnarray*}
p\left(\boldsymbol{Y}^{obs},\boldsymbol{Y}^{miss}|\boldsymbol{S},\boldsymbol{X},\boldsymbol{\Theta}\right) & = & \left(2\pi\right)^{-\frac{JT}{2}}\tau^{\frac{JT}{2}}\exp\left\{ -\frac{\tau}{2}\textrm{tr}\left(\boldsymbol{U}^{\top}\boldsymbol{U}\right)\right\} \\
 & = & \left(2\pi\right)^{-\frac{JT}{2}}\tau^{\frac{JT}{2}}\exp\left\{ -\frac{\tau}{2}\textrm{vec}\left(\boldsymbol{U}\right)^{\top}\textrm{vec}\left(\boldsymbol{U}\right)\right\} ,
\end{eqnarray*}
\[
\boldsymbol{U}=\boldsymbol{Y}-\boldsymbol{\Theta}-\boldsymbol{\Xi},
\]
where $\textrm{vec}\left(\cdot\right)$ denotes the column-wise vectorization
operator. The joint posterior distribution of the missing observations
of the responses and the unknown parameters is written as
\[
p\left(\boldsymbol{Y}^{miss},\boldsymbol{\Theta}|\boldsymbol{Y}^{obs},\boldsymbol{S},\boldsymbol{X}\right)\propto p\left(\boldsymbol{\Theta}\right)\prod_{\left(j,t\right)\in\mathcal{I}}p\left(y_{j,t}\left(0\right),y_{j,t}\left(1\right),\boldsymbol{S},\boldsymbol{X}|\boldsymbol{\Theta}\right),
\]
where $p\left(\boldsymbol{\Theta}\right)$ denotes the prior density
of $\boldsymbol{\Theta}$. Given $\boldsymbol{Y}^{obs}$ and $\boldsymbol{\Theta}$,
the conditional posterior distribution of the missing responses is
given by

\begin{eqnarray*}
p\left(\boldsymbol{Y}^{miss}|\boldsymbol{Y}^{obs},\boldsymbol{S},\boldsymbol{X},\boldsymbol{\Theta}\right) & \propto & \prod_{\left(j,t\right)\in\mathcal{I}}p\left(s_{j,t}|y_{j,t}\left(0\right),y_{j,t}\left(1\right),\boldsymbol{x}_{j,t},\boldsymbol{\Theta}\right)\\
 &  & \qquad\quad\times p\left(y_{j,t}\left(0\right),y_{j,t}\left(1\right)|\boldsymbol{x}_{j,t},\boldsymbol{\Theta}\right)p\left(\boldsymbol{x}_{j,t}|\boldsymbol{\Theta}\right).
\end{eqnarray*}
By the unconfoundedness assumption, the assignment mechanism $p\left(s_{j,t}|y_{j,t}\left(0\right),y_{j,t}\left(1\right),\boldsymbol{x}_{j,t},\boldsymbol{\Theta}\right)$
and the covariate distribution $p\left(\boldsymbol{x}_{j,t}|\boldsymbol{\Theta}\right)$
are ignorable:
\begin{eqnarray*}
p\left(\boldsymbol{Y}^{miss}|\boldsymbol{Y}^{obs},\boldsymbol{S},\boldsymbol{X},\boldsymbol{\Theta}\right) & \propto & \prod_{\left(j,t\right)\in\mathcal{I}_{1}}p\left(y_{j,t}\left(0\right),y_{j,t}\left(1\right)|\boldsymbol{x}_{j,t},\boldsymbol{\Theta}\right)\\
 &  & \times\prod_{\left(j,t\right)\in\mathcal{I}_{0}}p\left(y_{j,t}\left(0\right),y_{j,t}\left(1\right)|\boldsymbol{x}_{j,t},\boldsymbol{\Theta}\right)\\
 & \propto & \prod_{\left(j,t\right)\in\mathcal{I}_{1}}p\left(y_{j,t}\left(0\right)|y_{j,t}\left(1\right),\boldsymbol{x}_{j,t},\boldsymbol{\Theta}\right)\\
 &  & \times\prod_{\left(j,t\right)\in\mathcal{I}_{0}}p\left(y_{j,t}\left(1\right)|y_{j,t}\left(0\right),\boldsymbol{x}_{j,t},\boldsymbol{\Theta}\right).
\end{eqnarray*}
In turn, the conditional posterior of $\boldsymbol{\Theta}$ is proportional
to the product of the complete-data likelihood and the prior of $\boldsymbol{\Theta}$:
\[
p\left(\boldsymbol{\Theta}|\boldsymbol{Y}^{miss},\boldsymbol{Y}^{obs},\boldsymbol{S},\boldsymbol{X},\boldsymbol{\Theta}\right)\propto p\left(\boldsymbol{\Theta}\right)p\left(\boldsymbol{Y}^{obs},\boldsymbol{Y}^{miss}|\boldsymbol{S},\boldsymbol{X},\boldsymbol{\Theta}\right).
\]
We can conduct a posterior simulation using a Gibbs sampler: $\boldsymbol{Y}^{miss}$
and $\boldsymbol{\Theta}$ are alternately simulated from the corresponding
conditional posterior distributions. 

Once the approximation of the posterior distribution of $\boldsymbol{Y}^{obs}$
is obtained, we can evaluate treatment effects straightforwardly.
For instance, the posterior density of ATET can be represented as

\begin{eqnarray*}
p\left(ATET|\boldsymbol{Y}^{obs},\boldsymbol{X}\right) & = & \int\int\left[\frac{1}{\left|\mathcal{I}_{1}\right|}\sum_{\left(j,t\right)\in\mathcal{I}_{1}}\left(y_{j,t}\left(1\right)-y_{j,t}\left(0\right)\right)\right]\\
 &  & \quad\times p\left(\boldsymbol{Y}^{miss},\mathcal{\boldsymbol{\Theta}}|\boldsymbol{Y}^{obs},\boldsymbol{X}\right)p\left(\mathcal{\boldsymbol{\Theta}}\right)\textrm{d}\boldsymbol{Y}^{miss}\textrm{d}\boldsymbol{\Theta}.
\end{eqnarray*}
Given the posterior draws of the potential outcomes, the posterior
mean estimate of ATET is computed as

\[
\widehat{ATET}=\frac{1}{N_{post}}\sum_{i=1}^{N_{post}}\left[\frac{1}{\left|\mathcal{I}_{1}\right|}\sum_{\left(j,t\right)\in\mathcal{I}_{1}}\left(y_{j,t}\left(1\right)-y_{j,t}^{\left(i\right)}\left(0\right)\right)\right],
\]
where $y_{j,t}^{\left(i\right)}\left(0\right)$ denotes the $i$th
posterior draw of the potential outcome of unit $j$ at period $t$
and $N_{post}$ is the number of posterior draws used for the posterior
analysis. The posterior estimates of the variance/quantiles of the
posterior of $ATET$ are obtained analogously.

\subsection{Priors}

As the structure of the model indicates, unless some restrictions
are imposed, we cannot identify $\boldsymbol{\Gamma}$ and $\boldsymbol{Y}^{miss}$.
We induce $\boldsymbol{\Gamma}$ to be low rank and decompose it into
two parts as 
\[
\boldsymbol{\Gamma}=\boldsymbol{\Phi}\boldsymbol{\Psi}^{\top},\quad\text{with }\boldsymbol{\Phi}\in\mathbb{R}^{J\times H},\;\boldsymbol{\Psi}\in\mathbb{R}^{T\times H}.
\]
where $H<\min\left(J,T\right)$. Although this decomposition is not
unique, as $\boldsymbol{\Phi}$ and $\boldsymbol{\Psi}$ are not identified,
exact parameter identification is not necessary for our purpose:
we require the identification of the convolution, $\boldsymbol{\Gamma}$,
not that of its factorization, $\boldsymbol{\Phi}$ and $\boldsymbol{\Psi}$. 

Nevertheless, when $\boldsymbol{\Phi}$ and $\boldsymbol{\Psi}$ are
not identified, the posterior simulation can diverge, which is computationally
inefficient. We use a prior motivated by singular value decomposition
(SVD). When the SVD of $\boldsymbol{\Gamma}$ is represented as $\boldsymbol{\Gamma}=\boldsymbol{E}_{1}\boldsymbol{D}\boldsymbol{E}_{2}^{\top}$,
we interpret $\boldsymbol{\Psi}=\boldsymbol{E}_{2}$ as the right
orthonormal matrix and $\boldsymbol{\Phi}=\boldsymbol{E}_{1}\boldsymbol{D}$
as the product of the left orthonormal matrix $\boldsymbol{E}$ and
the diagonal matrix having the eigenvalues in its principal diagonal
$\boldsymbol{D}$. Two types of priors introduced in what follows
correspond to the interpretation of $\boldsymbol{\Phi}$ and $\boldsymbol{\Psi}$.

First, we restrict $\boldsymbol{\Psi}$ to be unitary, i.e., $\boldsymbol{\Psi}^{\top}\boldsymbol{\Psi}=\boldsymbol{I}_{H}$,
and assign a uniform Haar prior to $\boldsymbol{\Psi}$, $p\left(\boldsymbol{\Psi}\right)\propto\mathbb{I}\left(\boldsymbol{\Psi}\in\mathcal{M}_{T\times H}\right)$,
where $\mathcal{M}_{T\times H}$ denotes a Stiefel manifold with dimensions
of $T\times H$ and $\mathbb{I}\left(\cdot\right)$ denotes the indicator
function. This restriction implies that the covariance of the rows
of $\boldsymbol{\Psi}$ is $T\boldsymbol{I}_{H}$. Then, $\boldsymbol{\Gamma}=\left(\boldsymbol{\gamma}_{1},...,\boldsymbol{\gamma}_{T}\right)^{\top}$
can be regarded as being generated from a static factor model as 
\[
\boldsymbol{\gamma}_{t}=\boldsymbol{\Phi}\boldsymbol{\psi}_{t},\quad\boldsymbol{\psi}_{t}\sim\mathcal{N}\left(\boldsymbol{0}_{H},\;T\boldsymbol{I}_{H}\right),\quad t=1,...,T.
\]

Second, we arrange the relative magnitudes of the columns of $\boldsymbol{\Phi}$
in descending order. For this purpose, we adapt a cumulative shrinkage
process prior \cite{Legramanti2019} to our context. A prior of $\boldsymbol{\Phi}$
is specified by the following hierarchy: 
\begin{eqnarray*}
\phi_{j,h}|\lambda_{h} & \sim & \mathcal{N}\left(0,\lambda_{h}^{2}\right),\quad j=1,...,J;\;h=1,...,H,\\
\lambda_{h}|\pi_{h} & \sim & \pi_{h}\delta_{\lambda_{\infty}}+\left(1-\pi_{h}\right)\mathcal{IG}\left(\kappa_{1},\kappa_{2}\right),\quad h=1,...,H,\\
\pi_{h} & = & \sum_{l=1}^{h}\omega_{l},\quad\text{with }\omega_{l}=\zeta_{l}\prod_{m=1}^{l-1}\left(1-\zeta_{m}\right),\quad h=1,...,H,\\
\zeta_{h} & \sim & \mathcal{B}\left(1,\eta\right),\quad h=1,...,H-1,\\
\zeta_{H} & = & 1,
\end{eqnarray*}
where $\mathcal{IG}\left(a,b\right)$ is an inverse gamma distribution
with shape parameter $a$ and rate parameter $b$, and $\mathcal{B}\left(a,b\right)$
is a beta distribution (of the first kind) with scale parameters $a$
and $b$. The prior of $\phi_{j,h}$ is a scale mixture of normal
distributions. The prior distribution of the variances $\lambda_{h}$
belongs to a class of spike-and-slab priors (e.g., \cite{Ishwaran2005}),
in that the prior consists of spike $\delta_{\lambda_{\infty}}$ and
slab $\mathcal{IG}\left(\kappa_{1},\kappa_{2}\right)$. Although $\delta_{\lambda_{\infty}}$
can be zero, we set it to a small nonzero value for the ease of posterior
simulation \cite{Ishwaran2005,Legramanti2019}. The prior distribution
of the weights $\pi_{h}$ exploits the stick-breaking construction
of the Dirichlet process \cite{Ishwaran2001}. As $h$ grows, the
distribution of $\lambda_{h}$ concentrates around $\delta_{\lambda_{\infty}}$
since $\lim_{h\rightarrow\infty}\pi_{h}=1$ almost surely. 

In turn, for the remaining parameters, we employ standard priors.
For $\boldsymbol{\beta}$, we use an independent normal prior with
mean zero and precision $\alpha$, $\boldsymbol{\beta}\sim\mathcal{N}\left(\boldsymbol{0}_{L},\;\alpha^{-1}\boldsymbol{I}_{L}\right)$.
The prior distribution of $\tau$ is specified by a gamma distribution
with shape parameter $\nu_{1}$ and rate parameter $\nu_{2}$, $\tau\sim\mathcal{G}\left(\nu_{1},\nu_{2}\right)$. 

Although we do not consider them in this study, many alternative priors
can be used for $\boldsymbol{\Theta}$. Bhattacharya and Dunson \cite{Bhattacharya2011}
consider a prior similar to the cumulative shrinkage process prior,
called the multiplicative gamma process prior. This prior cannot simultaneously
control the rate of shrinkage and the prior for the active elements;
thus, it readily overshrinks the model. See \cite{Durante2017} and
\cite{Legramanti2019} for further discussion. In addition, many
fully Bayesian approaches exist for estimating or completing low-rank
matrices (e.g., \cite{Salakhutdinov2008,Ding2011}).
However, these approaches do not consider the parameter identification.
The only exception is Tang et al. \cite{Tang2019}. They factorize a possibly
rank-deficient matrix $\boldsymbol{\Gamma}$ into three parts as in
SVD, $\boldsymbol{\Gamma}=\boldsymbol{\Phi}\boldsymbol{D}\boldsymbol{\Psi}^{\top}$,
where $\boldsymbol{D}$ is diagonal. While they suppose $\boldsymbol{\Phi}$
and $\boldsymbol{\Psi}$ to be unitary as in this study, the diagonal
elements of $\boldsymbol{D}$ are not restricted: the ordering of
rows of $\boldsymbol{\Phi}$ and $\boldsymbol{\Psi}$ and the diagonal
elements of $\boldsymbol{D}$ are freely permuted along the posterior
simulation. 

\subsection{Posterior simulation}

For posterior simulation, we develop an MCMC sampler. We conduct posterior
simulations using a hybrid of two algorithms. To address the unitary
constraint, we sample $\boldsymbol{\Psi}$ using the geodesic Monte
Carlo on embedded manifolds \cite{Byrne2013}. As the conditionals
of the remaining parameters are standard, the remaining parameters
are updated via Gibbs steps. See the Appendix for the computational
details.

While Legramanti et al. \cite{Legramanti2019} adaptively tune the rank of a matrix
of interest, we prefix the rank of $\boldsymbol{\Gamma}$, $H$, for
several reasons. First, the unitary constraint on $\boldsymbol{\Psi}$
makes it difficult to change $H$ adaptively. Second, as our prior
pushes $\boldsymbol{\Gamma}$ to be low rank, it is unnecessary to
exactly specify the true rank of $\boldsymbol{\Gamma}$: if the $h^{\prime}$th
eigenvalue of $\boldsymbol{\Gamma}$ is negligible, the prior standard
deviation of the $h^{\prime}$th row of $\boldsymbol{\Phi}$ is inclined
to be $\delta_{\infty}$ (spike part). Therefore, we recommend choosing
a conservative value for $H$ or tuning $H$ based on test runs. 

\subsection{Extensions}

We mention some simple extensions. First, we can make the model more
robust to outliers by modeling the measurement errors using a distribution
with heavier tails than those of a normal distribution. For instance,
following \cite{Geweke1993}, the generalized Student's t error is modeled
as
\begin{eqnarray*}
u_{j,t}|\tau,\omega_{j,t} & \sim & \mathcal{N}\left(0,\;\tau^{-1}\omega_{j,t}^{-1}\right),\quad j=1,...,J;\;t=1,...,T,\\
\omega_{j,t}|\upsilon & \sim & \mathcal{G}\left(\frac{\upsilon}{2},\frac{\upsilon}{2}\right),\quad j=1,...,J;\;t=1,...,T,\\
\upsilon & \sim & f\left(\upsilon\right),
\end{eqnarray*}
where $\omega_{j,t}$ is an auxiliary random variable, $\upsilon$
is the degrees of freedom of $u_{j,t}$, and $f\left(\upsilon\right)$
is a prior distribution of $\upsilon$.

Second, to allow serial correlations in the error terms, their distribution
can be modeled as

\[
\boldsymbol{u}_{j}=\left(u_{j,1},...,u_{j,T}\right)^{\top}|\tau,\rho\sim\mathcal{N}\left(\boldsymbol{0}_{T},\;\tau^{-1}\boldsymbol{R}\right),
\]

\[
\boldsymbol{R}=\left(r_{t,t^{\prime}}\right),\;\text{with}\;r_{t,t^{\prime}}=\rho^{\left|t-t^{\prime}\right|},
\]
where $\boldsymbol{R}$ is a correlation matrix whose generic element
$r_{t,t^{\prime}}$ is specified as a function of an autocorrelation
parameter $\rho\in\left(-1,1\right)$. As the conditional posterior
of $\rho$ is not standard, $\rho$ is sampled using, e.g., the random-walk
Metropolis-Hastings algorithm. 

\subsection{Comparison with existing approaches}

The class of SCMs \cite{Abadie2003,Abadie2010} is closely related
to the proposed approach. In SCMs, ``synthetic'' untreated outcomes
are estimated as weighted sums of the untreated units. This approach
imposes three strong assumptions: no intercept, nonnegativity of the
weights, and weights that sum to one. However, none of these assumptions
appears plausible in many real cases. The proposed approach is free
from such restrictions. Doudchenko and Imbens \cite{Doudchenko2017}
propose an approach that does not impose any of these restrictions
on the weights and use a penalty similar to the elastic net estimator
\cite{Zou2005}. Amjad et al. \cite{Amjad2018} propose a robust synthetic control
method (RSCM). The difference between RSCM and the abovementioned
SCMs is that RSCM constructs a design matrix using the SVD of a matrix
composed of the outcomes of untreated units: SVD is used for dimension
reduction and denoising. Xu \cite{Xu2017} also considers a similar
modeling strategy.

All the existing non-Bayesian approaches, including \cite{Abadie2010},
\cite{Doudchenko2017}, \cite{Amjad2018}, and \cite{Xu2017} share
the same caveat: they cannot evaluate confidence intervals straightforwardly.
Abadie et al. \cite{Abadie2010} conduct a series of placebo studies, which can
be interpreted as permutation tests, to quantify the uncertainty of
an inference, but the size of the permutation tests may be distorted
as shown by \cite{Hahn2017}. No statistically sound method has been
developed to estimate confidence intervals of synthetic control methods.
Recently, Li \cite{Liforthcoming} proposes a subsampling method to
obtain confidence intervals. In contrast, our Bayesian approach can
estimate credible intervals as a byproduct of posterior simulation. 

Kim et al. \cite{Kim2019} develop a Bayesian version of Doudchenko and Imbens's
\cite{Doudchenko2017} approach. Instead of the elastic net penalty,
they propose the use of a shrinkage prior, e.g., the horseshoe prior
\cite{Carvalho2010}. As with Bayesian inference, their approach
can obtain credible intervals consistently. Our fully Bayesian approach
also enjoys the same advantage. Amjad et al. \cite{Amjad2018} also mention a
Bayesian version of RSCM, but the method is not fully Bayesian in
that the SVD of an outcome matrix is treated as given and uncertainty
about the decomposition is ignored.

Our proposal is closely related to Athey et al. \cite{Athey2018}, where an estimation
problem is treated as a matrix completion problem with a nuclear norm
penalty. Athey et al. \cite{Athey2018} call their estimator the
matrix completion with a nuclear norm minimization estimator (MC-NNM).
The prior of $\boldsymbol{\Gamma}$ used in our approach plays a similar
role to the nuclear norm penalty because the nuclear norm is a convex
relaxation of the rank constraint \cite{Fazel2001}. This family
of approaches involving matrix completion has two notable advantages
over SCMs. First, treatment is allowed to occur arbitrarily, not consecutively.
Second, while SCMs use only pretreatment observations for estimation,
this family exploits all the observations including the treated periods
(except treated outcomes). Therefore, this class is likely to be statistically
more efficient than SCMs, as shown in the simulation study below.
Similar to Amjad et al.'s \cite{Amjad2018} approach, the matrix
completion approaches intend to capture the underlying factor structure
of panel data. While in Amjad et al.'s \cite{Amjad2018} approach,
a threshold for truncating the eigenvalues of an outcome matrix must
be specified (hard thresholding), our approach and Amjad et al.'s
\cite{Amjad2018} approach do not because the cumulative shrinkage
process prior and the nuclear norm penalty automatically push the
model to be low rank (soft thresholding). As with other non-Bayesian
approaches, Athey et al.'s \cite{Athey2018} approach provides only
a point estimation, while our proposal readily estimates credible
intervals. 

Finally, Brodersen et al. \cite{Brodersen2015} and Ning et al. \cite{Ning2019} also develop
Bayesian approaches to causal inference that use structural time series
models, more specifically, state-space models. Brodersen et al.'s
\cite{Brodersen2015} approach relies on a univariate state-space
model, while Ning et al.'s \cite{Ning2019} approach uses a multivariate
state-space model that allows spatial correlations between units.
These two approaches are similar to ours in that both tend to obtain
potential outcomes using Bayesian methods. On the other hand, there
is a notable difference between their approaches and ours: their approaches
rely on the fit of a state-space model, while our approach exploits
the factor structure of panel data. As a consequence, our proposal
is better suited for typical panel data where due to the short sample,
it is difficult to recover the dynamics of the potential outcomes
from the observations.

\section{Application}

\subsection{Simulated data}

We conduct a simulation study to demonstrate the proposed approach.
In our experimental setting, only the $J$th unit is treated, and
it is exposed to the treatment during the last $T_{1}$ periods of
$T$. Let $T_{0}$ denote the number of untreated periods; thus, $T=T_{0}+T_{1}$.
The realized treated outcomes are specified by the sums of hypothetical
untreated outcomes $y_{J,t}\left(0\right)$ and the average treatment
effect on treated $ATET$:
\[
y_{J,t}\left(1\right)=y_{J,t}\left(0\right)+ATET,\quad t=T_{0}+1,...,T.
\]
 $y_{j,t}\left(0\right)$ is generated from a factor model: for $j=1,...,J$;
$t=1,...,T$,

\[
\left(y_{1,t}\left(0\right),...,y_{J,t}\left(0\right)\right)^{\top}=\boldsymbol{y}_{t}\left(0\right)=\boldsymbol{\Phi}\boldsymbol{\psi}_{t}+\boldsymbol{u}_{t},
\]
\[
\boldsymbol{\Psi}=\left(\boldsymbol{\psi}_{1},...,\boldsymbol{\psi}_{T}\right)^{\top},
\]
\[
\boldsymbol{u}_{t}=\left(u_{1,t},...,u_{J,t}\right)^{\top}\sim\mathcal{N}\left(\boldsymbol{0}_{J},\boldsymbol{I}_{J}\right).
\]
We do not include any covariates. 

We consider three types of data-generating processes (DGPs). The first
two types differ only in how $\boldsymbol{\Psi}$ is generated. In
the first case, called DGP-independent, latent factors are independently
distributed according to a normal distribution specified as 

\[
\boldsymbol{\psi}_{t}\sim\mathcal{N}\left(\boldsymbol{0}_{3},\boldsymbol{I}_{3}\right),\quad t=1,...T.
\]
The second case is called DGP-dependent, where the row of motion of
$\boldsymbol{\psi}_{t}=\left(\psi_{1,t},\psi_{2,t},\psi_{3,t}\right)^{\top}$
is specified by the following process:
\[
\psi_{j,1}=\epsilon_{j,1},\quad j=1,2,3,
\]
\[
\begin{cases}
\psi_{1,t}=0.6\psi_{1,t-1}+\epsilon_{1,t}\\
\psi_{2,t}=0.4\psi_{2,t-1}+\epsilon_{2,t}, & t=2,3,...,T,\\
\psi_{3,t}=0.2\psi_{3,t-1}+\epsilon_{3,t}
\end{cases}
\]
\[
\boldsymbol{\epsilon}_{t}=\left(\epsilon_{1,t},\epsilon_{2,t},\epsilon_{3,t}\right)^{\top}\sim\mathcal{N}\left(\boldsymbol{0}_{3},\boldsymbol{I}_{3}\right),\quad t=1,...,T.
\]
Entries in the factor loading$\boldsymbol{\Phi}$ are generated independently
from a standard normal distribution:

\[
\boldsymbol{\Phi}=\left(\phi_{j,t}\right),\quad\text{with}\;\phi_{j,t}\sim\mathcal{N}\left(0,1\right).
\]
The third case, DGP-weighted, is motivated by a simulation study in
\cite{Kim2019}. In this setting, outcomes of untreated units are
generated from a multivariate normal distribution, and outcomes of
treated units are constructed as weighted sums of untreated units:
\[
y_{J,t}=\sum_{j=1}^{J-1}\alpha_{j}y_{j,t}+u_{J,t},\quad u_{J,t}\sim\mathcal{N}\left(0,\sigma^{2}\right),\quad t=1,...,T,
\]
\[
\boldsymbol{y}_{1:J-1,t}\left(0\right)\sim\mathcal{N}\left(\boldsymbol{\mu},\boldsymbol{\Sigma}\right),\quad t=1,...,T,
\]

\[
\boldsymbol{\alpha}=\left(\alpha_{j}\right),\quad\alpha_{j}=\begin{cases}
3 & j=1\\
2 & j=2\\
1 & j=3\\
0 & j=4,...,J-1
\end{cases},
\]
\[
\boldsymbol{\Sigma}=\left(\sigma_{i,j}\right),\quad\sigma_{i,j}=\begin{cases}
10 & i=j\\
0.5 & i\neq j
\end{cases}.
\]
$\boldsymbol{\mu}$ is specified as follows: 

\[
\boldsymbol{\mu}=\left(\mu_{j}\right),\quad,\mu_{j}=\begin{cases}
10 & j=1\\
20 & j=2\\
30 & j=3\\
40 & j=4\\
15 & j=5,...,9
\end{cases},\;\text{for }J=10,
\]
and
\[
\boldsymbol{\mu}=\left(\mu_{j}\right),\quad,\mu_{j}=\begin{cases}
10 & j=1\\
20 & j=2\\
30 & j=3\\
40 & j=4\\
15 & j=5,...,10\\
25 & j=11,...,20\\
35 & j=21,...,30\\
45 & j=31,...,39
\end{cases},\;\text{for }J=40.
\]

We compare six alternative approaches.
\begin{enumerate}
\item The first is the original synthetic control method described in \cite{Abadie2010}
(SCM-ABD).
\item The second is a method proposed by \cite{Doudchenko2017} (SCM-DI). 
\item The third is a Bayesian approach developed by Kim et al. \cite{Kim2019}. According
to their simulation study, specifications with the horseshoe \cite{Carvalho2010}
and spike-and-slab \cite{Ishwaran2005} priors outperform other alternatives.
While the performances of these two priors are comparable, posterior
simulation using the horseshoe prior is faster. Thus, we consider
the horseshoe prior for Kim et al.'s \cite{Kim2019} approach and
refer to this specific approach as BSCM. We sample weighting parameters
in the observation model using the elliptical slice sampler \cite{Hahn2019}
and obtain the remaining parameters (noise variance and shrinkage
parameters) using a Gibbs sampler, as in \cite{Makalic2016}. 
\item The fourth is the robust synthetic control method introduced by \cite{Amjad2018}
(RSCM). Specifically, we consider their primary choice described in
Algorithm 1 of the original paper (p. 8). 
\item The fifth is the matrix completion with a nuclear norm minimization
estimator \cite{Athey2018} (MC-NNM). 
\item The seventh is the proposed approach, Bayesian matrix completion with
cumulative shrinkage process prior (BMC-CSP). The prefixed hyperparameters
for the cumulative shrinkage process prior are chosen following \cite{Legramanti2019}
as $\eta=5$ and $\kappa_{1}=\kappa_{2}=2$. While Legramanti et al. \cite{Legramanti2019}
use $\delta_{\infty}=0.05$, we use a smaller value, $\delta_{\infty}=0.01$.
The maximum rank of $\boldsymbol{\Theta}$ is set to $H=\textrm{min}\left(J,T\right)$,
where $\left\lceil \right\rceil $ denotes the ceiling function. 
\end{enumerate}
To ensure a fair comparison, we use the same prior for the error variance
in BSCM as in the proposed approach. We choose the hyperparameters
as $\nu_{1}=\nu_{2}=0.001$, inducing the prior of $\tau$ to be fairly
noninformative. For MC-NNM, we choose tuning parameters via five-fold
cross-validation, where the training samples are randomly chosen without
replacement. For SCM-DI and RSCM, the tuning parameters are determined
by forward chaining: the tuning parameters are chosen by minimizing
the mean squared errors of one-step-ahead out-of-sample predictions,
and the training sample is initially set to five and expanded sequentially
to $T_{0}-1$. For BSCM and BMC-CSP, we obtain 40,000 draws after
discarding the initial 10,000. All the posterior simulations pass
Geweke's \cite{Geweke1992} convergence test at a significance level of 5\%.

We consider four types of sample size, namely, combinations of $J\in\left\{ 5,20\right\} $
and $T_{0}\in\left\{ 10,40\right\} $, and the length of the treated
periods is fixed to $T_{1}=20$. A total of 200 experiments are conducted
for each case. As noted earlier, an estimation of ATE amounts to an
estimation of potential outcomes. Therefore, we evaluate the alternatives
based on the precision of the estimates of $y_{j,t}\left(0\right)$,
$t=T_{0}+1,...,T$, measured by the mean of the sum of the squared
errors (MSE) and the mean of the sum of absolute errors (MAE). For
the Bayesian approaches, we compute posterior means of predicted potential
outcomes. We also report the mean computation time measured in seconds
(Time).\footnote{We wrote all the programs in Matlab R2019b (64 bit) and executed them
on an Ubuntu Desktop 18.04 LTS (64 bit), running on AMD Ryzen Threadripper
1950X (4.2GHz).} For each experiment, the MSE and MAE are normalized by the corresponding
values for SCM-ABD, (i.e., the smaller the values are, the better).

Table 1 summarizes the results of the simulation study for DGP-independent.
In terms of MSE and MAE, irrespective of the combination of $\left(J,T_{0}\right)$,
the proposed approach consistently outperforms the others. The recently
proposed alternatives are comparable with or worse than SCM-ABD. In
this setting, RSCM is consistently inferior to the original SCM-ABD.
In terms of computational time, as expected, the Bayesian approaches
are slower than the non-Bayesian options. Indeed, BMC-CSP is computationally
heavy, but the computational cost is not prohibitive. In our simulation
study, SCM-DI is slow to converge, possibly due to nonsmooth objective
functions. The simulation results for DGP-dependent are reported in
Table 2. In terms of MSE and MAE, RSCM, MC-NNM, and BMC-SCP perform
best. Table 3 summarizes the results for DGP-weighted. SCM-DI and
BSCM perform very well because this DGP is exactly consistent with
the DGPs of the models. In contrast to the other DGPs, the predictive
accuracy of RSCM and MC-NNM is much worse than that of the others,
including SCM-ABD. Although BMC-SCP performs worse than SCM-DI and
BSCM, it consistently outperforms the remaining approaches. In summary,
while the relative finite-sample performance of the alternative approaches
depends on the DGP, the proposed approach, BMC-SCP, is fairly competitive
under various circumstances.

\begin{table}
\caption{Results of simulation study (1): DGP-independent}
\label{tab:1}
\begin{tabular}{clr@{\extracolsep{0pt}.}lr@{\extracolsep{0pt}.}lr@{\extracolsep{0pt}.}l}
\hline 
$\left(J,T_{0}\right)$ & Approach & \multicolumn{2}{c}{MSE} & \multicolumn{2}{c}{MAE} & \multicolumn{2}{c}{Time}\tabularnewline
\hline 
\multirow{6}{*}{$\left(5,10\right)$} & (1) SCM-ABD & 1&00 & 1&00 & 0&1\tabularnewline
 & (2) SCM-DI & 1&48 & 1&17 & 24&7\tabularnewline
 & (3) BSCM & 1&60 & 1&23 & 11&3\tabularnewline
 & (4) RSCM & 1&22 & 1&08 & 6&5\tabularnewline
 & (5) MC-NNM & 0&99 & 0&98 & 0&8\tabularnewline
 & (6) BMC-CSP & 0&95 & 0&96 & 92&7\tabularnewline
\hline 
\multirow{6}{*}{$\left(5,40\right)$} & (1) SCM-ABD & 1&00 & 1&00 & <0&1\tabularnewline
 & (2) SCM-DI & 1&54 & 1&21 & 96&0\tabularnewline
 & (3) BSCM & 1&53 & 1&21 & 28&3\tabularnewline
 & (4) RSCM & 1&27 & 1&11 & 8&1\tabularnewline
 & (5) MC-NNM & 1&09 & 1&04 & 1&6\tabularnewline
 & (6) BMC-CSP & 0&90 & 0&95 & 507&8\tabularnewline
\hline 
\multirow{6}{*}{$\left(20,10\right)$} & (1) SCM-ABD & 1&00 & 1&00 & <0&1\tabularnewline
 & (2) SCM-DI & 0&96 & 0&97 & 175&5\tabularnewline
 & (3) BSCM & 0&97 & 0&97 & 14&9\tabularnewline
 & (4) RSCM & 1&29 & 1&10 & 53&2\tabularnewline
 & (5) MC-NNM & 1&00 & 0&98 & 0&8\tabularnewline
 & (6) BMC-CSP & 0&90 & 0&93 & 96&1\tabularnewline
\hline 
\multirow{6}{*}{$\left(20,40\right)$} & (1) SCM-ABD & 1&00 & 1&00 & <0&1\tabularnewline
 & (2) SCM-DI & 1&16 & 1&06 & 728&6\tabularnewline
 & (3) BSCM & 1&68 & 1&27 & 46&1\tabularnewline
 & (4) RSCM & 1&59 & 1&24 & 74&3\tabularnewline
 & (5) MC-NNM & 1&09 & 1&04 & 2&1\tabularnewline
 & (6) BMC-CSP & 0&94 & 0&97 & 514&5\tabularnewline
\hline 
\end{tabular}
\end{table}

\begin{table}
\caption{Results of simulation study (2): DGP-dependent}
\label{tab:2}
\begin{tabular}{clr@{\extracolsep{0pt}.}lr@{\extracolsep{0pt}.}lr@{\extracolsep{0pt}.}l}
\hline 
$\left(J,T_{0}\right)$ & Approach & \multicolumn{2}{c}{MSE} & \multicolumn{2}{c}{MAE} & \multicolumn{2}{c}{Time}\tabularnewline
\hline 
\multirow{6}{*}{$\left(5,10\right)$} & (1) SCM-ABD & 1&00 & 1&00 & <0&1\tabularnewline
 & (2) SCM-DI & 1&44 & 1&15 & 23&9\tabularnewline
 & (3) BSCM & 1&57 & 1&21 & 10&3\tabularnewline
 & (4) RSCM & 0&71 & 0&84 & 6&1\tabularnewline
 & (5) MC-NNM & 0&73 & 0&85 & 0&6\tabularnewline
 & (6) BMC-CSP & 0&71 & 0&84 & 93&8\tabularnewline
\hline 
\multirow{6}{*}{$\left(5,40\right)$} & (1) SCM-ABD & 1&00 & 1&00 & <0&1\tabularnewline
 & (2) SCM-DI & 1&53 & 1&22 & 97&9\tabularnewline
 & (3) BSCM & 1&42 & 1&18 & 26&0\tabularnewline
 & (4) RSCM & 0&81 & 0&89 & 7&6\tabularnewline
 & (5) MC-NNM & 0&88 & 0&93 & 1&1\tabularnewline
 & (6) BMC-CSP & 0&79 & 0&89 & 508&7\tabularnewline
\hline 
\multirow{6}{*}{$\left(20,10\right)$} & (1) SCM-ABD & 1&00 & 1&00 & <0&1\tabularnewline
 & (2) SCM-DI & 0&86 & 0&92 & 170&6\tabularnewline
 & (3) BSCM & 0&87 & 0&92 & 14&0\tabularnewline
 & (4) RSCM & 0&75 & 0&86 & 52&0\tabularnewline
 & (5) MC-NNM & 0&76 & 0&86 & 0&5\tabularnewline
 & (6) BMC-CSP & 0&75 & 0&86 & 97&0\tabularnewline
\hline 
\multirow{6}{*}{$\left(20,40\right)$} & (1) SCM-ABD & 1&00 & 1&00 & <0&1\tabularnewline
 & (2) SCM-DI & 1&47 & 1&19 & 698&4\tabularnewline
 & (3) BSCM & 1&75 & 1&30 & 42&3\tabularnewline
 & (4) RSCM & 0&87 & 0&93 & 65&9\tabularnewline
 & (5) MC-NNM & 0&90 & 0&95 & 1&5\tabularnewline
 & (6) BMC-CSP & 0&88 & 0&94 & 518&6\tabularnewline
\hline 
\end{tabular}
\end{table}

\begin{table}
\caption{Results of simulation study (3): DGP-weighted}
\label{tab:3}
\begin{tabular}{clr@{\extracolsep{0pt}.}lr@{\extracolsep{0pt}.}lr@{\extracolsep{0pt}.}l}
\hline 
$\left(J,T_{0}\right)$ & Approach & \multicolumn{2}{c}{MSE} & \multicolumn{2}{c}{MAE} & \multicolumn{2}{c}{Time}\tabularnewline
\hline 
\multirow{6}{*}{$\left(5,10\right)$} & (1) SCM-ABD & 1&00 & 1&00 & <0&1\tabularnewline
 & (2) SCM-DI & 0&03 & 0&17 & 25&4\tabularnewline
 & (3) BSCM & 0&03 & 0&17 & 17&3\tabularnewline
 & (4) RSCM & 23&28 & 5&70 & 12&7\tabularnewline
 & (5) MC-NNM & 15&18 & 4&62 & 3&2\tabularnewline
 & (6) BMC-CSP & 0&63 & 0&78 & 90&4\tabularnewline
\hline 
\multirow{6}{*}{$\left(5,40\right)$} & (1) SCM-ABD & 1&00 & 1&00 & <0&1\tabularnewline
 & (2) SCM-DI & 0&39 & 0&60 & 70&1\tabularnewline
 & (3) BSCM & 0&43 & 0&63 & 54&1\tabularnewline
 & (4) RSCM & 14&67 & 4&60 & 23&9\tabularnewline
 & (5) MC-NNM & 13&88 & 4&44 & 6&3\tabularnewline
 & (6) BMC-CSP & 0&71 & 0&83 & 507&3\tabularnewline
\hline 
\multirow{6}{*}{$\left(20,10\right)$} & (1) SCM-ABD & 1&00 & 1&00 & <0&1\tabularnewline
 & (2) SCM-DI & 0&02 & 0&13 & 211&2\tabularnewline
 & (3) BSCM & 0&02 & 0&13 & 21&0\tabularnewline
 & (4) RSCM & 26&56 & 6&17 & 86&7\tabularnewline
 & (5) MC-NNM & 11&74 & 3&96 & 3&5\tabularnewline
 & (6) BMC-CSP & 0&05 & 0&19 & 92&2\tabularnewline
\hline 
\multirow{6}{*}{$\left(20,40\right)$} & (1) SCM-ABD & 1&00 & 1&00 & <0&1\tabularnewline
 & (2) SCM-DI & 0&03 & 0&17 & 622&1\tabularnewline
 & (3) BSCM & 0&03 & 0&17 & 84&1\tabularnewline
 & (4) RSCM & 22&77 & 5&74 & 185&6\tabularnewline
 & (5) MC-NNM & 9&70 & 3&63 & 7&4\tabularnewline
 & (6) BMC-CSP & 0&41 & 0&63 & 511&9\tabularnewline
\hline 
\end{tabular}
\end{table}

\subsection{Real data}

As an illustration, we apply the proposed approach to evaluate California's
tobacco control program implemented in 1988. We replicate Abadiet
et al.'s \cite{Abadie2010} study using the same data, annual state-level
panel data spanning 1970 to 2000.\footnote{The data and the Matlab program were downloaded from Jens Hainmueller's
personal website. (https://web.stanford.edu/\textasciitilde{}jhain/synthpage.html)} The first 19 years are the pretreatment period. Only California is
treated, while the other 38 states are used as control units. We include
seven time-invariant covariates: log of gross domestic product per
capita, percentage share of 15\textendash 24\textendash year\textendash old
people in the population, retail price, beer consumption per capita,
and cigarette sales per capita in 1980 and 1975; see \cite{Abadie2010}
for further details. We use the same hyperparameters as in the simulation
study. We draw 100,000 posterior samples and use the last 80,000 samples
for posterior analysis.\footnote{We also conduct a posterior simulation where the unitary constraint
on $\boldsymbol{\Psi}$ is removed and $\boldsymbol{\Psi}$ is sampled
via a standard Gibbs step, but this approach is unsuccessful because
the Markov chains diverge, resulting in numerical error.} 

Figure 1 compares the realized per capita cigarette sales in California
(solid black line), the potential per capita cigarette sales in ``synthetic
California'' obtained using the original SCM \cite{Abadie2010}
(dashed black line), and the posterior mean estimates of the corresponding
potential outcomes obtained by the proposed method (solid red line).
The estimates obtained using the proposed method are in line with
the estimates obtained using the original SCM. Posterior estimates
of 90\% and 70\% credible sets are also reported (shaded areas). As
the credible sets do not include the realized California, the program
has statistically significant effects on tobacco consumption in California,
confirming the conclusion in the original paper. 

\begin{figure}
  \scalebox{.75}{
     \includegraphics{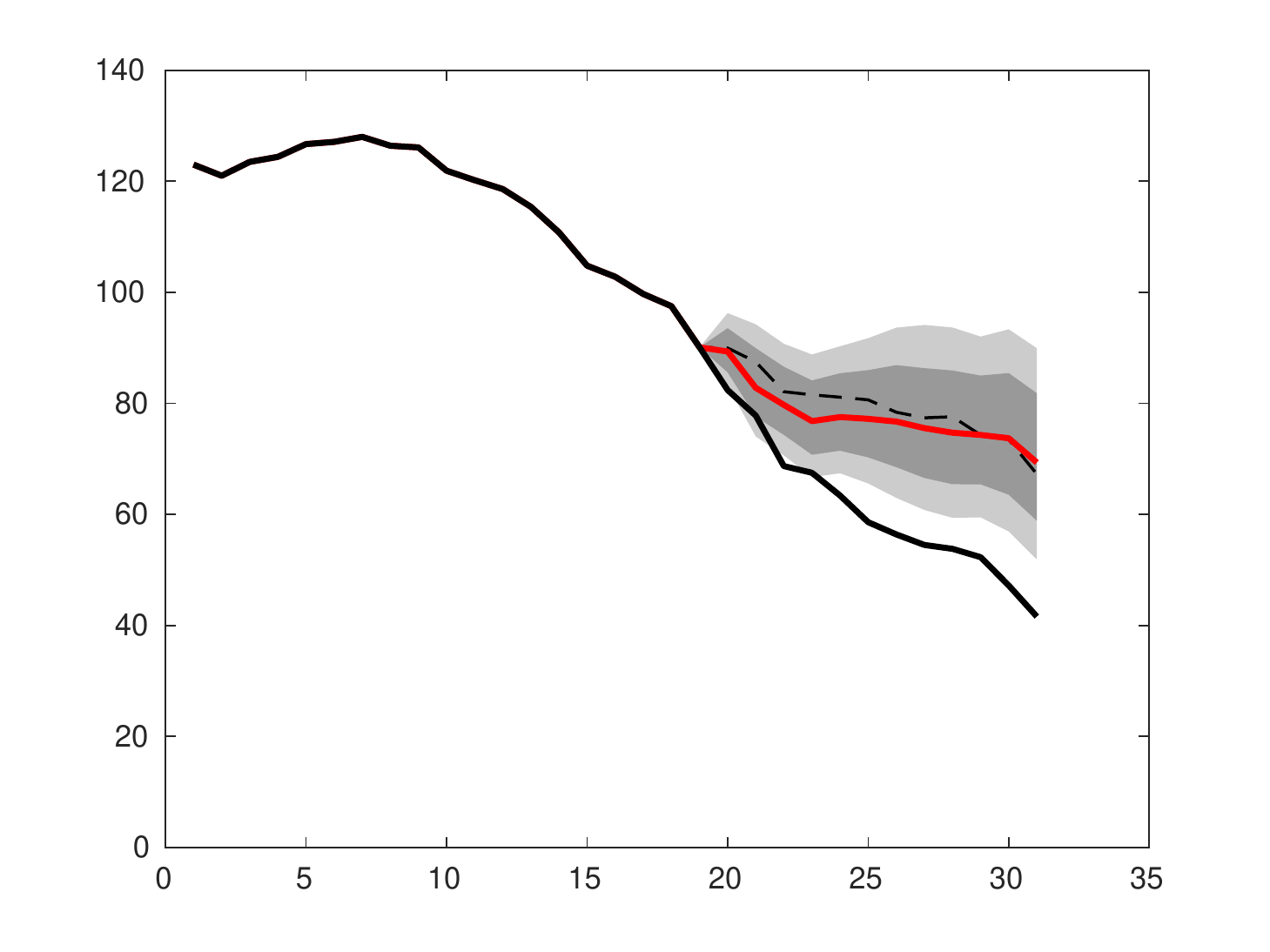}
  }
  \caption{The solid black line traces the realized per capita cigarette
sales in California. The dashed black line and the solid red line
trace the estimated potential per capita cigarette sales using SCM-ABD
and BMC-CSP, respectively. The light and dark shaded areas indicate
the 90\% and 70\% credible sets, respectively.
}
\label{fig:1}
\end{figure}

Figure 2 depicts the posterior estimates of some rows of $\boldsymbol{\Phi}$,
which can be interpreted as state-specific loadings. While the estimates
have different patterns, reflecting heterogeneity in US states, their
magnitude is roughly decreasing with $h$ as intended by the prior.
Figure 3 plots the posterior mean estimates of the eigenvalues of
$\boldsymbol{\Gamma}$, which suggests that approximately half of
the eigenvalues are not essential. 

\begin{figure}
  \includegraphics{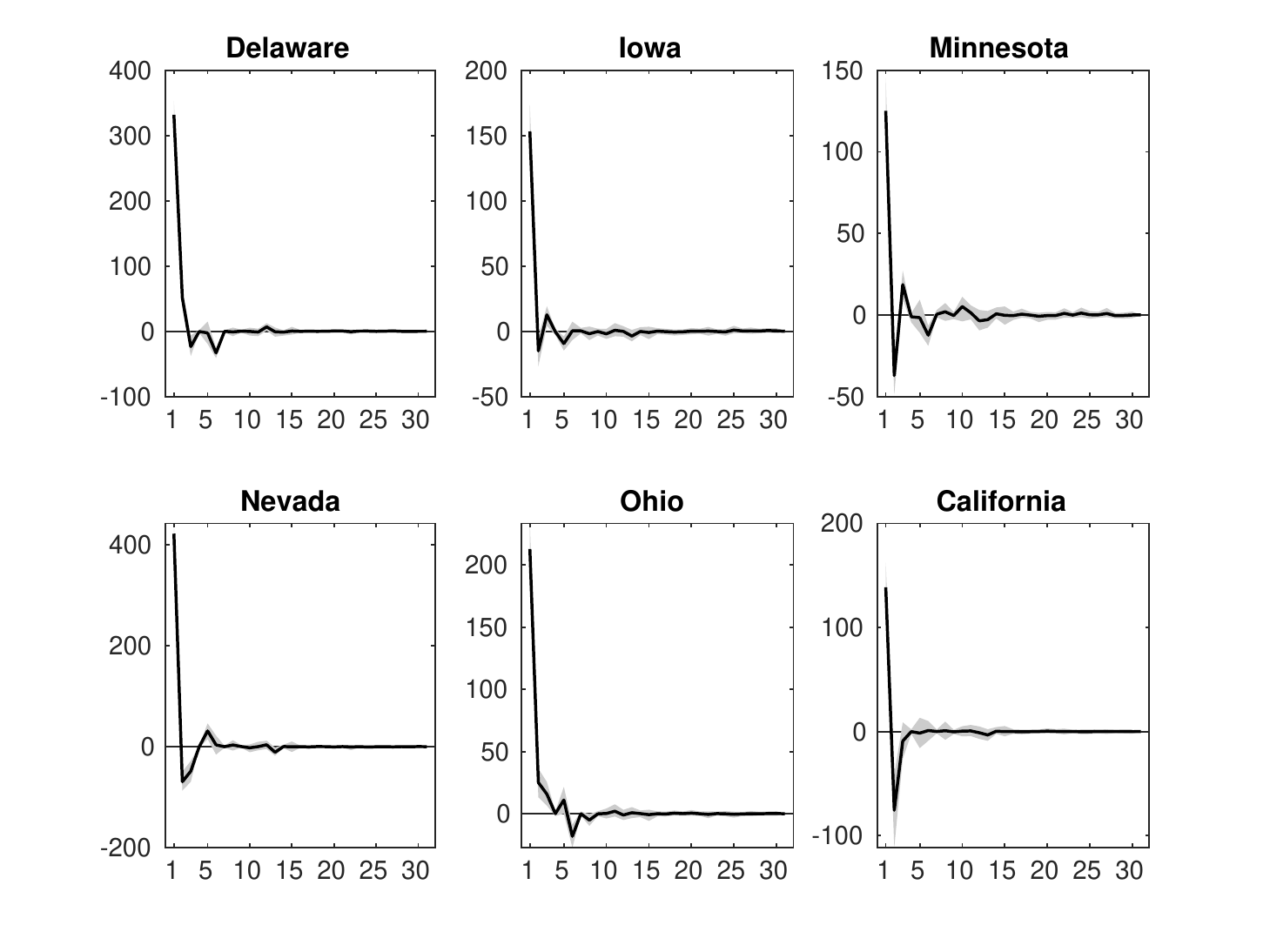}
  \caption{The solid black lines trace the posterior mean estimates of
the rows of $\boldsymbol{\Phi}$. The shaded areas indicate the 90\%
credible sets.
}
\label{fig:2}
\end{figure}

\begin{figure}
  \scalebox{.75}{
    \includegraphics{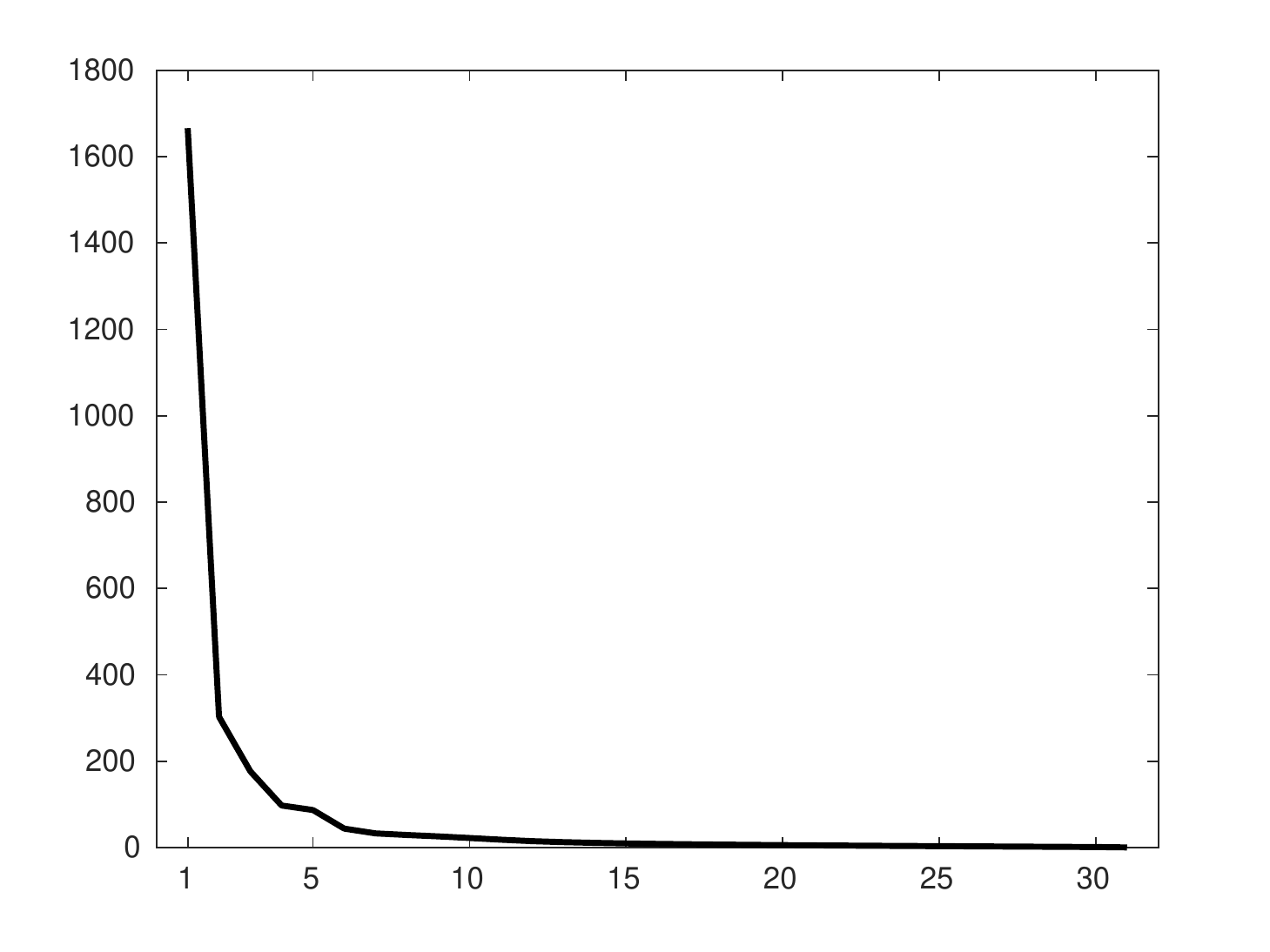}
}
  \caption{The solid black lines trace the posterior mean estimates of
the eigenvalues of $\boldsymbol{\Gamma}$.
}
\label{fig:3}
\end{figure}

\section{Concluding Remarks}

This study develops a novel Bayesian approach to causal analysis using
panel data. We treat the problem of inferring a treatment effect as
a matrix completion problem: counterfactual untreated outcomes that
are actually treated are inferred using a data augmentation technique.
We also propose a prior structured to help identification and to obtain
a low-rank approximation of the panel data. In contrast to existing
non-Bayesian methods, the proposed Bayesian approach can estimate
credible intervals straightforwardly. By means of a series of simulation
studies, we show that the proposed approach outperforms the existing
ones in terms of the prediction of hypothetical untreated outcomes,
that is, the accuracy of the treatment effect estimates. 

While asymptotic argument is not absolutely necessary for Bayesian analysis, 
there is a need to investigate frequentist (asymptotic) properties of the 
proposed approach, such as posterior consistency and Bernstein-von-Mises 
theorem. However, up to the author's knowledge, there is no published work 
on frequentist properties of Bayesian matrix factorization/completion, except 
\cite{Mai2015}.\footnote{Mai and Alquier \cite{Mai2015} propose a Bayesian 
estimator for the matrix completion method and provide an oracle inequality 
for this estimator. However, they employ a uniform prior, and the proof critically 
depends on this prior choice; thus, their discussion is not easily extended 
to other environments.} The author hopes that this paper stimulates further 
theoretical studies in the related research horizons.

\section*{Appendix: Computational Details}

This appendix describes the computational details of the posterior
simulation of the proposed approach. The joint posterior is specified
as 
\begin{eqnarray*}
p\left(\boldsymbol{Y}^{miss},\boldsymbol{\Phi},\boldsymbol{\Psi},\boldsymbol{\beta},\tau,\boldsymbol{\zeta},\boldsymbol{\Lambda}|\boldsymbol{Y}^{obs},\boldsymbol{X}\right) & \propto & p\left(\boldsymbol{Y}^{obs}|\boldsymbol{Y}^{miss},\boldsymbol{\Phi},\boldsymbol{\Psi},\boldsymbol{\beta},\tau;\boldsymbol{X}\right)p\left(\boldsymbol{Y}^{miss}\right)p\left(\tau\right)\\
 &  & \times p\left(\boldsymbol{\beta}\right)p\left(\boldsymbol{\Psi}\right)p\left(\boldsymbol{\Phi}|\boldsymbol{\Lambda}\right)p\left(\boldsymbol{\Lambda}|\boldsymbol{\zeta}\right)p\left(\boldsymbol{\zeta}\right)\\
 & \propto & \tau^{\frac{JT}{2}}\exp\left\{ -\frac{\tau}{2}\textrm{tr}\left(\boldsymbol{U}^{\top}\boldsymbol{U}\right)\right\} \times\tau^{\nu_{1}-1}\exp\left(-\nu_{2}\tau\right)\\
 &  & \times\exp\left\{ -\frac{\alpha}{2}\boldsymbol{\beta}^{\top}\boldsymbol{\beta}\right\} \times\mathbb{I}\left(\boldsymbol{\Psi}\in\mathcal{M}_{T\times H}\right)\\
 &  & \times\prod_{j=1}^{J}\exp\left\{ -\frac{1}{2}\boldsymbol{\phi}_{\left(j\right)}^{\top}\textrm{diag}\left(\lambda_{1}^{-1},...,\lambda_{H-1}^{-1}\right)\boldsymbol{\phi}_{\left(j\right)}\right\} \\
 &  & \times\prod_{h=1}^{H}\left[\left(\sum_{l=1}^{h}\zeta_{l}\prod_{m=1}^{l-1}\left(1-\zeta_{m}\right)\right)\delta_{\lambda_{\infty}}\right.\\
 &  & +\left(1-\sum_{l=1}^{h}\zeta_{l}\prod_{m=1}^{l-1}\left(1-\zeta_{m}\right)\right)\\
 &  & \left.\times\left\{ \frac{\kappa_{2}^{\kappa_{1}}}{\Gamma\left(\kappa_{1}\right)}\lambda_{h}^{-\kappa_{1}-1}\exp\left(-\frac{\kappa_{2}}{\lambda_{h}}\right)\right\} \right]\\
 &  & \times\prod_{h=1}^{H-1}\left(1-\zeta_{h}\right)^{\eta-1},
\end{eqnarray*}
where $\boldsymbol{\zeta}=\left(\zeta_{1},...,\zeta_{H-1}\right)^{\top}$
and $\boldsymbol{\Lambda}=\textrm{diag}\left(\lambda_{1},...,\lambda_{H}\right)$.
Each sampling block is specified in what  follows.

\paragraph{Sampling $\boldsymbol{\Phi}$}

Each row of $\boldsymbol{\Phi}$ is sampled from a multivariate normal
distribution. For $j=1,...,J$,

\[
\boldsymbol{\phi}_{\left(j\right)}|\text{rest}\sim\mathcal{N}\left(\boldsymbol{m}_{\boldsymbol{\phi}_{\left(j\right)}},\boldsymbol{P}_{\boldsymbol{\phi}_{\left(j\right)}}^{-1}\right),
\]
\[
\boldsymbol{m}_{\boldsymbol{\phi}_{\left(j\right)}}=\boldsymbol{P}_{\boldsymbol{\phi}_{\left(j\right)}}^{-1}\boldsymbol{\Psi}^{\top}\left(\boldsymbol{y}_{\left(j\right)}-\boldsymbol{\xi}_{\left(j\right)}\right),
\]
\[
\boldsymbol{P}_{\boldsymbol{\phi}_{\left(j\right)}}=\boldsymbol{\Lambda}^{-1}+\boldsymbol{\Psi}^{\top}\boldsymbol{\Psi},
\]
\[
\boldsymbol{Y}=\left(\boldsymbol{y}_{\left(1\right)},...,\boldsymbol{y}_{\left(J\right)}\right)^{\top},\quad\boldsymbol{\Xi}=\left(\boldsymbol{\xi}_{\left(1\right)},...,\boldsymbol{\xi}_{\left(J\right)}\right)^{\top}.
\]

\paragraph{Sampling the shrinkage parameters}

The conditional of $z_{h}$ is specified as 

\[
p\left(z_{h}=l|\text{rest}\right)\propto\begin{cases}
\omega_{l}\text{N}\left(\boldsymbol{\phi}_{h}|\boldsymbol{0}_{J},\;\lambda_{\infty}\boldsymbol{I}_{J}\right), & l=1,...,h,\\
\omega_{l}\text{t}_{2\kappa_{1}}\left(\boldsymbol{\phi}_{h}|\boldsymbol{0}_{J},\;\frac{\kappa_{2}}{\kappa_{1}}\boldsymbol{I}_{J}\right), & l=h+1,...,H,
\end{cases}
\]
where $\textrm{N}\left(\boldsymbol{x}|\boldsymbol{a},\boldsymbol{B}\right)$
is the PDF of a multivariate normal distribution with mean $\boldsymbol{a}$
and covariance $\boldsymbol{B}$ evaluated at $\boldsymbol{x}$ and
$\textrm{t}_{c}\left(\boldsymbol{x}|\boldsymbol{a},\boldsymbol{B}\right)$
is the PDF of a multivariate t distribution with location parameter
$\boldsymbol{a}$, scale parameter $\boldsymbol{B}$, and $c$ degrees
of freedom. The sampling distributions of $\zeta_{l}$ and $\lambda_{h}$
are 
\begin{eqnarray*}
\zeta_{h}|\text{rest} & \sim & \mathcal{B}\left(1+\sum_{l=1}^{H}\mathbb{I}\left(z_{l}=h\right),\quad\eta+\sum_{l=1}^{H}\mathbb{I}\left(z_{l}>h\right)\right),\quad h=1,...,H-1,\\
\lambda_{h}|\text{rest} & \sim & \mathbb{I}\left(z_{h}\leq h\right)\delta_{\theta_{\infty}}+\left(1-\mathbb{I}\left(z_{h}\leq h\right)\right)\mathcal{IG}\left(\kappa_{1}+\frac{J}{2},\;\kappa_{2}+\frac{1}{2}\sum_{j=1}^{J}\phi_{j,h}^{2}\right),\quad h=1,...,H.
\end{eqnarray*}

\paragraph{Sampling $\boldsymbol{\Psi}$}

To sample $\boldsymbol{\Psi}$, we employ the geodesic Monte Carlo
on embedded manifolds developed by \cite{Byrne2013}. The algorithm
for sampling $\boldsymbol{\Psi}$ is summarized in Algorithm 1. Let
$\pi\left(\boldsymbol{\Psi}\right)$ be the posterior density of $\boldsymbol{\Psi}$
conditional on the other parameters. Then the gradient with respect
to $\boldsymbol{\Psi}$ is derived as
\[
\nabla_{\boldsymbol{\Psi}}\log\pi\left(\boldsymbol{\Psi}\right)=\tau\left(\boldsymbol{Y}-\boldsymbol{\Xi}\right)^{\top}\boldsymbol{\Phi}-\tau\boldsymbol{\Psi}\boldsymbol{\Phi}^{\top}\boldsymbol{\Phi}.
\]
The step size $\varepsilon$ is adaptively tuned to maintain the average
acceptance rate near a target value $a^{*}$. In the $i$th iteration,
$\varepsilon$ is updated as follows:
\[
\log\left(\varepsilon\right)\leftarrow\log\left(\varepsilon\right)+i^{-1/\varsigma}\left(a^{*}-\bar{a}_{i}\right),
\]
where $\bar{a}_{i}$ is the average acceptance rate in the $i$th
iteration and $\varsigma\in\left(0.5,1\right)$ is a tuning parameter.
We choose $a^{*}=0.6$ and $\varsigma=0.6$.\footnote{The target acceptance rate $a^{*}$ is chosen based on a multivariate
effective sample size \cite{Vats2019}.} The number of steps is fixed to five, $N_{step}=5$.

\paragraph{Sampling $\boldsymbol{\beta}$}

$\boldsymbol{\beta}$ is simulated from a multivariate normal distribution:

\[
\boldsymbol{\beta}|\text{rest}\sim\mathcal{N}\left(\boldsymbol{m}_{\boldsymbol{\beta}},\boldsymbol{P}_{\boldsymbol{\beta}}^{-1}\right),
\]
\[
\boldsymbol{m}_{\boldsymbol{\beta}}=\tau\boldsymbol{P}_{\boldsymbol{\beta}}^{-1}\boldsymbol{X}^{\top}\textrm{vec}\left(\boldsymbol{Y}-\boldsymbol{\Theta}\right),
\]
\[
\boldsymbol{P}_{\boldsymbol{\beta}}=\alpha\boldsymbol{I}_{L}+\tau\boldsymbol{X}^{\top}\boldsymbol{X},
\]
\[
\boldsymbol{X}=\left(\begin{array}{c}
\boldsymbol{X}_{1}\\
\vdots\\
\boldsymbol{X}_{J}
\end{array}\right),\;\text{with}\;\boldsymbol{X}_{j}=\left(\begin{array}{c}
\boldsymbol{x}_{j,1}^{\top}\\
\vdots\\
\boldsymbol{x}_{j,T}^{\top}
\end{array}\right).
\]

\paragraph{Sampling $\tau$}

$\tau$ is updated via the following gamma distribution:

\[
\tau|\text{rest}\sim\mathcal{G}\left(\nu_{1}+\frac{JT}{2},\;\nu_{2}+\frac{1}{2}\textrm{tr}\left(\boldsymbol{U}^{\top}\boldsymbol{U}\right)\right).
\]

\paragraph{Sampling $\boldsymbol{Y}^{miss}$}

The conditional posterior distribution of a missing observation of
unit $j$ in time period $t$ is a normal distribution,
\[
y_{j,t}\left(0\right)|\text{rest}\sim\mathcal{N}\left(\gamma_{j,t}+\xi_{j,t},\;\tau^{-1}\right),\quad\left(j,t\right)\in\mathcal{I}_{1}.
\]

%
%

\begin{acknowledgements}
The author would like to thank Hideki Konishi, Yasuhiro Omori, and
Hisatoshi Tanaka for their helpful discussions.
\end{acknowledgements}

%
\section*{Conflict of interest}
The author declares that there is no conflict of interest.

\bibliographystyle{spmpsci}      
\bibliography{reference}

%
%

\end{document}